\documentclass[11pt]{article}

\usepackage{hyperref}
\usepackage{amsmath,amsfonts,amssymb}
\hypersetup{dvips,dvipdfm,colorlinks=true,urlcolor=magenta,filecolor=magenta,linktoc=page,citecolor=red,linkcolor=blue,bookmarks=true}
\usepackage{graphicx,epstopdf}
\usepackage{multirow,array}
\begin{document}
	\title{Exploring Traversable Wormholes in $f(Q)$ gravity: Shadows and Quasinormal modes}
	\author{Madhukrishna Chakraborty$^1$ \footnote{\url{ chakmadhu1997@gmail.com} (corresponding author)} ~~and ~~Subenoy Chakraborty$^2$ \footnote{\url{schakraborty.math@gmail.com}}}
	\date{%
		\small	$^1$Department of Mathematics, Techno India University, Kolkata-700091, West Bengal, India\\%
		$^2$Department of Mathematics, Brainware University, Kolkata-700125, West Bengal, India\\
		$^2$ Shinawatra University, Thailand\\
		$^2$ INTI International University, Malaysia\\[2ex]%
		{}
	}
	\maketitle
	\begin{abstract}
		The present work deals with some WH solutions in $f(Q)$ gravity theory for non-constant red-shift function subject to a power law model of $f(Q)$, $Q$ being the non-metricity scalar. Important properties of a WH configuration like the asymptotic flatness, traversability, flairing out condition and energy conditions have been checked for the obtained solutions. Shadows for these WHs have been determined and effect of non-metricity has been discussed. Finally, the paper gives an essence of Quasinormal modes and grey body factors for traversable WHs  in general and $f(Q)$ wormholes in particular indicating a distinct signature of non-metricity influencing these observational tools.
	\end{abstract}
	\section{Introduction}
	Wormholes (WH) are tunnel-like hypothetical astrophysical objects \cite{flamm}. The geometry of the WHs is a very peculiar yet interesting in the sense that it connects two disconnected or distant regions in such a way that singularity is removed. The topology of WH is non-simply connected and is thus fundamentally different from the topology of ordinary flat space-time. Actually, this non-simply connectedness is a property that allows for paths between regions that can not be continuously deformed into one another. The interesting thing is that such type of geometric structure can be obtained as a solution of Einstein's field equations. WHs were originally found by Einstein and Rosen \cite{Einstein:1935tc} which is why they are also called Einstein-Rosen bridge \cite{Einstein:1935tc}. WHs can be categorized into traversable WHs (TWH) and non-traversable WHs based on whether a test particle can pass through them. Einstein-Rosen bridge is a non traversable one and hence it is considered as nothing but a mathematical artifact. On the other hand, TWHs exhibit both mathematical as well as physical characteristics and hence of more interest. Ellis \cite{ellis} obtained a novel WH solution which is a spherically symmetric solution of Einstein's field equations with a ghost massless scalar field. Later, the Ellis WHs were proved traversable by Morris and Throne \cite{Morris:1988cz}. The term WH was first coined by Misner and Wheeler \cite{Wheeler:1957mu}. In General relativity (GR), violation of energy conditions (ECs) is necessary for formation of a WH and matter violating the ECs is exotic in nature. Also, from the point of view of observation exotic matter has been accepted as the successful candidate responsible for the late time acceleration. This paved the way for the study of phantom WH solutions. If WHs do exist, many unusual phenomena can be expected like interstellar, intergalactic or interuniverse trips and even time-travel/ time machine. This motivates the researchers to work on TWHs to get insights regarding some unknown universal phenomena.
	
	Modified theories of gravity are well accepted alternatives to Dark Energy (DE) to explain the accelerating cosmic expansion. So, it is interesting to test the viable WH solutions in extended theories of gravity. In literature, many studies in this context have been carried out like in Rastall gravity \cite{rastall}, $f(R)$ gravity \cite{fr}, $4D$ Einstein Gauss Bonnet Gravity \cite{egb},  Einstein Cartan theory \cite{ec} etc. Interestingly, it is found that exotic matter is not expected in many cases or even if it is required then the quantity of exotic matter can be minimized by using the notion of ``volume integral quantifier" which quantifies the total amount of matter violating ECs. This motivates us to study WH configuration in a recently popular theory of gravity namely, $f(Q)$ gravity. This gravity theory was proposed by Jimenez et. al. \cite{jib}. In $f(Q)$ gravity, both torsion and curvature vanish and role of gravity comes via the non-metricity $Q$. This means that the gravitational interaction is encoded in the non-metricity scalar $Q$, leading to second-order field equations as in GR but there is a scope of generalization via arbitrary functions $f(Q)$. The fundamental difference between teleparallel gravity and GR lies in the role of affine metric connection. The pathway to form new modified gravity theory from ``symmetric teleparallel gravity" is via the non-metricity scalar $Q$ and its extension to a function $f(Q)$ in the Lagrangian giving rise to $f(Q)$ modified gravity theory. The basis of this theory is the generalization of Riemannian geometry described by Weyl geometry. Working with $f(Q)$ gravity theory has got a couple of motivations. First of all, this theory supports various observations like Cosmic Microwave Background Radiation (CMBR), Supernova type Ia, Baryonic Acoustic Oscillations (BAO) \cite{bao}, growth data \cite{gd}, red-shift space distortion (RSD) \cite{rsd} etc. Studies reveal that $f(Q)$ gravity may challenge the standard $\Lambda CDM$ scenario \cite{cdm}. Also, this theory passes big bang nucleosynthesis (BBN) constraints \cite{bbn}. Further, ECs and Newtonian limit have been studied in $f(Q)$ gravity theory \cite{nl}. Concept of static and spherically symmetric BHs have been studied in \cite{static}. Lin and Zhai have investigated the application of $f(Q)$ gravity to the spherically symmetric configuration \cite{lin}. Wang et. al. have studied static and spherically symmetric solutions with an anisotropic fluid for general $f(Q)$ model \cite{wang}. Sharma et. al. have explored the WH solutions in the background of symmetric teleparallel gravity \cite{sharma} to have matter satisfying ECs to support WH formation. Hassan et.al. have also studied TWHs in $f(Q)$ gravity by considering two specific equation of state (EoS) \cite{sahoo}. They have presented solutions for a specific shape function for linear form of $f(Q)$ and found that the presented solutions in \cite{sah} violate the ECs. Banerjee et.al. \cite{banerjee} used a constant red-shift function with different shape functions to construct WH solutions with some known $f(Q)$ functions. They have found solutions which have not obeyed the ECs. The study \cite{Singh:2025ufl} uses a specific ansatz within $f(Q)$ gravity theory to examine the characteristics of spherically symmetric anisotropic compact stars and aims at enhancing comprehension of these typical entities by examining the physical properties of the compact star model using the Durgapal geometry within the context of 
	gravity.
	
	In GR, violation of NEC is required in order to sustain TWHs hinting at the presence of exotic matter. The motivation to study TWHs in $f(Q)$ gravity is too see whether the geometric contribution from $f(Q)$ can lead to violation of NEC even without exotic matter. The geometry of $f(Q)$ gravity is expected to provide new classes of solutions, including regular wormholes with better physical viability. Unlike $f(R),~f(Q)$ avoids higher order derivatives and is less pathological than other modified theories of gravity like $f(R),~f(T)$ etc.
	
	Investigating shadows cast by compact objects has emerged as a very promising area of research in the recent times. For Black-holes (BHs), studying shadows is an observational signature that gives insights to BH physics. Bardeen originally discussed shadows \cite{bardeen} and his work attracted profound interest in literature. Therefore, a natural expectation regarding the study of similar WH shadows has emerged lately. Nedkova, Tinchevand Yazadjev \cite{ned} studied shadows of rotating WH. They calculated the boundary of the shadow in the presence of an extended source behind the WH. Observational evidence for shadows of compact objects at the core of $SgrA*$ and $M87$ have motivated to search for various astrophysical models. Investigating shadows of WH play a pivotal role in formulating valuable data on its structure, including throat size, spin and potential accretion processes. Shadows of a variety of compact objects like BH, WH, Gravastars etc have been covered in \cite{sh}.

	On the other hand, Quasinormal modes (QNMs) and grey body factors are very crucial concepts in the arena of wave dynamics in curved space-time. They describe how perturbations such as Gravitational waves (GWs) behave near a compact astrophysical object. QNMs are the characteristic oscillations of a BH or other compact objects that result from perturbations. Grey body factors describe how incoming radiation from  BH is partially transmitted and partially reflected due to the curvature of space-time. The correspondence between QNMs and grey body factors has been established for spherically symmetric asymptotically flat or de-Sitter BHs \cite{qnms}. Grey body factors might exhibit greater stability against deformations of the near horizon geometry \cite{gb} as compared to the overtones of QNMs which are highly sensitive to such deformations. The correspondence between QNMs and grey body factors for BHs have been found using WKB approach. Given that the boundary conditions for QNMs and grey body factors are the same for BHs and WHs, it is reasonable to consider such similar correspondence in case of WHs too. While, the literature on QNMs of WHs is quite vivid, the calculation of grey body factors has been carried out for specific cases.

	 $f(Q)$ models show promising perturbation stability. The scalar, electromagnetic and gravitational perturbations around the WHs in $f(Q)$ gravity lead to modified Regge-Wheeler like equations due to extra geometric terms. This affects QNMs, changing ringdown spectrum,  potentially observable in gravitational wave data. QNMs might act like fingerprints of the gravity theory. QNMs in WH configuration could indicate non-Riemannian geometry like that in $f(Q)$. On the other hand, grey body factors depend on effective potential barrier seen by outgoing radiation. Modified space-time structure in $f(Q)$ can lead to distinct transmission/ reflection coefficients, thereby potentially altering observable fluxes from WH throat. Distinct grey body spectra provide additional observational handles to differentiate $f(Q)$ WHs from GR BHs or WHs in other theories.
	
	Although, the analysis of WH shadows and associated QNMs in modified gravity frameworks such as $f(Q)$ gravity offer promising insights for unveiling observational signatures beyond GR, it is essential to approach these with undue care. Recent studies have revealed that certain formulations of $f(Q)$ theories may exhibit strong coupling pathologies particularly around maximally symmetric backgrounds like Minkowski space-time \cite{Saha:2025cfs}, \cite{Capozziello:2024mxh}, \cite{Gomes:2023tur}. These issues stem from the loss of propagating degrees of freedom in the linearized regime, leading to an ill-defined perturbative expansion some cases. Nevertheless, such pathologies are often highly model dependent and may be circumvented by choosing well behaved functional forms of $f(Q)$ or by restricting to non-Minkowskian or dynamically curved space-time such as those found in WH space-times. The pathology may be avoided around more general geometry and by constraining model parameters. Moreover, the perturbative stability analysis via QNMs and the behavior of the grey body factors provide an indirect probe of such strong coupling effects, making their inclusion in this study both timely and necessary.
	
	In this work, we present some WH solutions for non trivial and non constant red shift function. Moreover, we check the ECs, asymptotic flatness, traversability and flairing out conditions in these solutions subject to a power law model of $f(Q)$ gravity. Consequently, both the effective potential in the context of shadows and QNMs have been obtained in the backdrop of power law $f(Q)$ model. Finally, we give an essence of QNMs and grey body factors for the present model under consideration. In all cases, we explicitly study the effect of gravity that comes via non-metricity $Q$.
	The layout of the paper is as follows: Section 2 deals with the basics of WH configuration and $f(Q)$ gravity. Section 3 deals with the WH solutions in $f(Q)$ gravity and the study of ECs. Shadows of $f(Q)$ WHs have been investigated in Section 4. Section 5 discusses the general idea of QNMs and grey body factors in the context of TWHs. Section 6 gives the QNM spectra and grey body factors for the models under consideration. Finally the paper ends with concluding remarks in Section 7.
	\section{Basics of WH configuration and $f(Q)$ gravity}
	\subsection{A brief overview of $f(Q)$ gravity}
	The action for symmetric teleparallel gravity is given by \cite{sahoo2}
	\begin{equation}
		\mathcal{S}=\int \dfrac{1}{2}f(Q)\sqrt{-g}~d^{4}x+\int \mathcal{L}_{m}\sqrt{-g}~d^{4}x\label{eq1}
	\end{equation}
	where, $f(Q)$ is a generic function of the non metricity scalar $Q$, $g=\det(g_{\mu\nu})$ and $\mathcal{L}_{m}$ is the matter Lagrangian density. The non-metricity tensor and its trace can be defined as follows \cite{Chakraborty:2024sco}, \cite{q}:
	\begin{eqnarray}
		Q_{\lambda\mu\nu}=\nabla_{\lambda}g_{\mu\nu}\\
		Q_{\alpha}=Q_{\alpha~~\mu}^{\mu}~,~\tilde{Q_{\alpha}}=Q_{\alpha\mu}^{\mu}
	\end{eqnarray}
	With the help of non-metricity we can write the superpotential as 
	\begin{eqnarray}
		P_{~\mu\nu}^{\alpha}=\dfrac{1}{4}\left(-Q_{~\mu\nu}^{\alpha}+2Q^{~~\alpha}_{(\mu~~\nu)}+Q^{\alpha}g_{\mu\nu}-\tilde{Q^{\alpha}}g_{\mu\nu}-\delta^{\alpha}_{(\mu^{Q}_{~~\nu})}\right)
	\end{eqnarray} where the trace of the non-metricity tensor takes the form 
	\begin{equation}
		Q=-Q_{\alpha\mu\nu}P^{\alpha\mu\nu}
	\end{equation} and is known as the non-metricity scalar. By the definition, the energy momentum tensor for the fluid content of the space-time can be expressed as 
	\begin{equation}
		T_{\mu\nu}=\dfrac{-2}{\sqrt{-g}}\dfrac{\delta(\sqrt{-g}\mathcal{L}_{m})}{\delta g^{\mu\nu}}
	\end{equation}
	By varying the action (\ref{eq1}) w.r.t metric tensor $g_{\mu\nu}$ we get,
	\begin{equation}
		T_{\mu\nu}=\dfrac{-2}{\sqrt{-g}}\nabla_{\gamma}(\sqrt{-g}f_{Q}P_{\mu\nu}^{\gamma})-\dfrac{1}{2}g_{\mu\nu}f-f_{Q}\left(P_{\mu\gamma i}Q_{\nu}^{\gamma i}-2Q_{\gamma i\mu}P_{\nu}^{\gamma i}\right)
	\end{equation}
	and
	\begin{equation}
		\nabla_{\mu}\nabla_{\nu}\left(\sqrt{-g}f_{Q}P_{\mu\nu}^{\gamma}\right)=0
	\end{equation} where $f_{Q}=\dfrac{df(Q)}{dQ}$.
	\subsection{An outline of WH configuration and ECs}
	Line element of a general spherically symmetric static WH space-time of the Morris Thorne class (traversable) is given by \cite{Mehdizadeh:2015jra}-\cite{Halder:2019urh}
	\begin{equation}
		ds^{2}=-e^{2\Phi(r)}dt^{2}+\dfrac{1}{\left(1-\dfrac{b(r)}{r}\right)}dr^{2}+r^{2}d\Omega_{2}^{2}\label{eq9}
	\end{equation}
	where $d\Omega_{2}^{2}=d\theta^{2}+\sin^{2}\theta d\phi^{2}$, $\Phi(r)$ is called the red-shift function and it is a function of the radial coordinate $r$ such that $r_{0}\leq r<\infty$. $r_{0}$ is the radius of the WH throat. Value of $e^{2\Phi(r)}$ is always finite everywhere to avoid horizon or singularity. The function $\Phi(r)$ is used to detect the red-shift of the signal by a distant observer and gives information about the radial tidal force. $b(r)$ is called the shape function that determines the shape of the WH. There are certain restrictions on $b(r)$. They are listed as follows:
	\begin{enumerate}
		\item $b(r_{0})=r_{0}$ is the throat condition and $b(r)<r$ for $r>r_{0}$.
		\item Flairing out condition: $\dfrac{b(r)-rb'(r)}{b^{2}(r)}>0$ where $b'(r)=\dfrac{db(r)}{dr}$.
		\item Asymptotic flatness: $\dfrac{b(r)}{r}\rightarrow 0$ as $r\rightarrow \infty$.
	\end{enumerate} 
	The ECs that play a crucial role in formation of a feasible WH are given by \cite{mc}
	\begin{enumerate}
		\item Null Energy Condition (NEC): $\rho+p_{r}\geq0$, $\rho+p_{t}\geq0$.
		\item Weak Energy Condition (WEC): $\rho\geq0$,  $\rho+p_{r}\geq0$, $\rho+p_{t}\geq0$.
		\item Strong Energy Condition (SEC):  $\rho+p_{r}\geq0$, $\rho+p_{t}\geq0$, $\rho+p_{r}+2p_{t}\geq0$.
		\item Dominant Energy Condition (DEC): $\rho-|p_{r}|\geq0$, $\rho-|p_{t}|\geq0$.
	\end{enumerate}
	Actually, the constraint of minimum radius at the throat with traversability criteria impose huge tension at the throat. This is further balanced by the violation of NEC at the throat i.e, $\rho+p_{r}<0$ i.e, $\dfrac{b(r)-rb'(r)}{b^{2}(r)}>0$ which is essentially the criteria for WH traversability. Raychaudhuri equation (RE) plays a very important role in understanding traversability of WHs and their connection to the violation of ECs in GR. RE describes evolution of a congruence of null geodesics and is given by \cite{Chakraborty:2024khs}
	\begin{equation}
		\dfrac{d\Theta}{d\lambda}=-\dfrac{\Theta^{2}}{2}-2\sigma^{2}+2\omega^{2}-R_{\mu\nu}k^{\mu}k^{\nu}
	\end{equation} where $\Theta$ is called the expansion scalar, $\sigma$ is called the anisotropy scalar, $\omega$ is the vorticity scalar and $\lambda$ is the affine parameter. Focusing theorem (FT) is the most important consequence of RE and it states that if $R_{\mu\nu}k^{\mu}k^{\nu}\geq 0$, then the congruence focus within some finite value of the affine parameter. Further, Landau \cite{Kar:2006ms} showed in his book \textit{Classical Theory of Fields} that defocusing avoids singularity. Now, $R_{\mu\nu}k^{\mu}k^{\nu}<0$ is the criteria for defocusing and it is the condition for violation of ECs that leads to traversability in WHs. Therefore, violation of ECs, defocusing of geodesic, no horizon/ singularity and traversability are all equivalent. This can be explained in the following manner:
	
	In order to have TWHs it is required to violate NEC. RE is key to this. For null geodesic congruence, the RE simplifies and shows that if NEC is violated then $\Theta$ decreases leading to geodesic focusing/ convergence (i.e, no WH throat). However, it is interesting to note that topology of WH is such that in order to avoid the convergence/ singularity two simply connected regions are amalgamated by removing the singularity via a feasible WH throat. Thus, to allow a WH throat where geodesics do not focus the term $R_{\mu\nu}k^{\mu}k^{^{\nu}}<0$ for null vectors $k^{\mu}$ must hold somewhere implying violation of NEC. Hence RE \cite{Kar:2006ms} provides an insight of geometrical justification for needing exotic matter (violating NEC) in wormhole space-times.
	
	 Thus, RE is critical in wormholes for 
		\begin{enumerate}
			\item predicting the existence of a WH throat via geodesic divergence.
			\item demonstrating the requirement for exotic matter (via violation of NEC)
			\item connecting geometry of space-time with physical energy matter content as in GR
			\item providing valuable insights to how bundle of light rays/ photons behave when they traverse through or near a WH which is the foundation of gravitational lensing and subsequently formation of WH shadows.
		\end{enumerate}
	
	\section{WH solutions in $f(Q)$ gravity: Traversability and Energy Conditions}
	In order to find the WH solutions in $f(Q)$ gravity, we consider the WH metric given by (\ref{eq9}). The expression for non-metricity scalar $Q$ is given by
	\begin{equation}
		Q=\dfrac{-2}{r}\left(1-\dfrac{b(r)}{r}\right)\left(2\Phi'(r)+\dfrac{1}{r}\right)\label{eq11}
	\end{equation} 
	The field equations can be written as  \cite{sahoo}, \cite{sah}, \cite{sahoo2}
	\scriptsize
	\begin{eqnarray}	\rho=\dfrac{f_{Q}}{r}\left(\dfrac{1}{r}-\dfrac{(rb'+b)}{r^{2}}+2\Phi'(r)\left(1-\dfrac{b(r)}{r}\right)\right)+\dfrac{2f'_{Q}}{r}\left(1-\dfrac{b(r)}{r}\right)+\dfrac{f(Q)}{2}\label{eq12}\\
		p_{r}=-f_{Q}\left(\dfrac{2}{r}\left(1-\dfrac{b(r)}{r}\right)\left(2\Phi'(r)+\dfrac{1}{r}\right)-\dfrac{1}{r^{2}}\right)-\dfrac{f(Q)}{2}\label{eq13}\\
		p_{t}=-\dfrac{f_{Q}}{r}\left(\left(1-\dfrac{b(r)}{r}\right)\left(\dfrac{1}{r}+\Phi'(r)(3+r\Phi'(r))+r\Phi''(r)\right)-\dfrac{(rb'(r)-b(r))}{2r^{2}}(1+r\Phi'(r))\right)-\dfrac{1}{r}\left(1-\dfrac{b(r)}{r}\right)(1+r\Phi'(r))f'_{Q}-\dfrac{f(Q)}{2}\label{eq14}
	\end{eqnarray}
	\normalsize
	It is to be noted that there are three equations namely equations (\ref{eq12}), (\ref{eq13}) and (\ref{eq14}) containing six unknowns namely, $b(r),~\Phi(r),~\rho,~p_{r},~p_{t}$ and $f(Q)$. So, we assume some mathematical conditions to obtain solution.
	\subsection{Solution I} 
	To simplify the calculations and find a suitable red-shift function, we choose the following constraint
	\begin{equation}
		3r\Phi'(r)+r^{2}\Phi'(r)^{2}+r^{2}\Phi''(r)=0
	\end{equation}
	The above constraint yields,
	\begin{equation}
		\Phi(r)=\ln \left(c_{2}-\dfrac{c_{1}}{r^{2}}\right)\label{eq16}
	\end{equation} where $c_{1},~c_{2}$ are arbitrary constants. It is to be noted that, $\Phi$ is a smooth function of $r$ in the region $r\geq r_{0}$ and $e^{\Phi(r)}\rightarrow c_{2}$ as $r\rightarrow \infty$. Thus, there is no horizon in $r\geq r_{0}$.
	
\begin{figure}[h!]
		\begin{minipage}{0.3\textwidth}
			\centering\includegraphics[height=5cm,width=5.5cm]{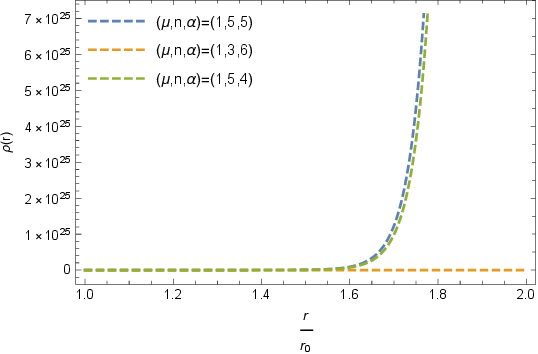}
		\end{minipage}~~~~~~~
		\begin{minipage}{0.3\textwidth}
			\centering\includegraphics[height=5cm,width=5.5cm]{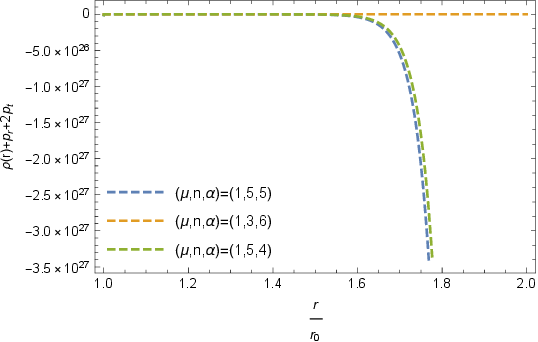}
		\end{minipage}\hfill
		\begin{minipage}{0.3\textwidth}
			\centering\includegraphics[height=5cm,width=5.5cm]{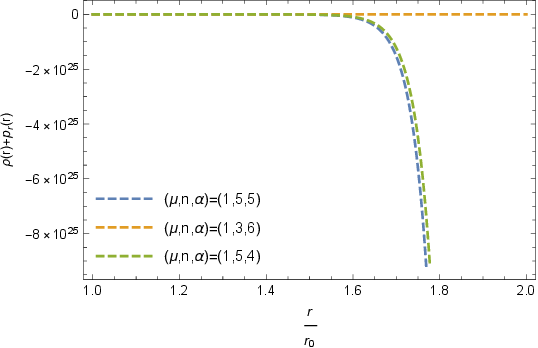}
		\end{minipage}
		\begin{minipage}{0.3\textwidth}
			\centering\includegraphics[height=5cm,width=5.5cm]{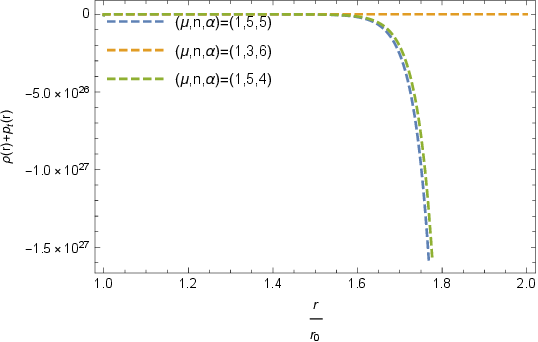}
		\end{minipage}
		\begin{minipage}{0.3\textwidth}
			\centering\includegraphics[height=5cm,width=5.5cm]{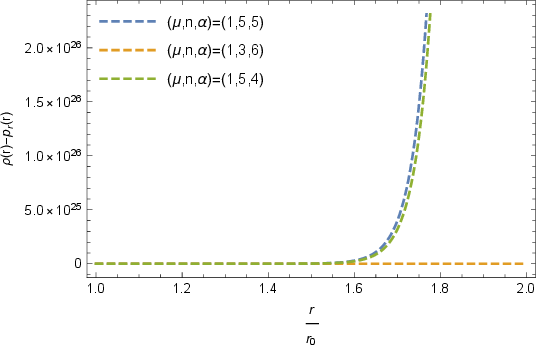}
		\end{minipage}
		\begin{minipage}{0.3\textwidth}
			\centering\includegraphics[height=5cm,width=5.5cm]{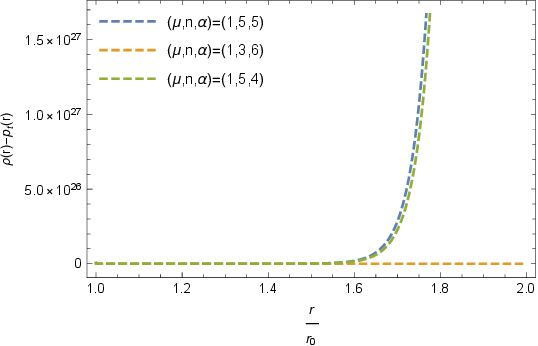}
		\end{minipage}
		\begin{minipage}{0.85\textwidth}\caption{NEC,WEC,SEC and DEC in case of Solution I for various choices of the model parameters specified in each panel.}\label{f1}
		\end{minipage}
	\end{figure}

	To determine $b(r)$, we consider the following constraint
	\begin{equation}
		\dfrac{1-\dfrac{b(r)}{r}}{r}-\dfrac{(rb'(r)-b(r))}{4r^{2}}=\mu r^{m}
	\end{equation}
	The above differential equation can be solved to get
	\begin{equation}
		b(r)=r-\left(\dfrac{4\mu}{m+5}\right)r^{m+2}+\dfrac{b_{0}}{r^{3}}
	\end{equation}
	Criteria for asymptotic flatness i.e, $\dfrac{b(r)}{r}\rightarrow 0$ as $r\rightarrow\infty$ gives $m=-1$ and $\mu=1$. Thus, the WH shape function is given by 
	\begin{equation}
		b(r)=\dfrac{b_{0}}{r^{3}}\label{eq19}
	\end{equation} where $b_{0}=b(r_{0})$.
	One can easily find the radius of the WH throat $r_{0}$ using the relation $b(r_{0})=r_{0}$ and it comes out to be $r_{0}=1$. Using equations (\ref{eq11}), (\ref{eq16}) and (\ref{eq19}), we get
	\begin{equation}
		Q=-\dfrac{2}{r}\left(1-\dfrac{1}{r^{4}}\right)\left(\dfrac{\dfrac{c_{1}}{r^{3}}+\dfrac{c_{2}}{r}}{c_{2}-\dfrac{c_{1}}{r^{2}}}\right)\label{eq20*}
	\end{equation}  
	With the solution $b(r)=\dfrac{1}{r^{3}}$, it is found that $rb'(r)-b(r)<0$ for all $r\geq r_{0}$. Thus, flairing out criteria holds true. Now, we choose the power law model for $f(Q)$ as $f(Q)=\alpha Q^{n}$ where, $\alpha,~n$ are the model parameters. The motivation behind choosing the power law model of $f(Q)$ comes from several physical and mathematical considerations \cite{Chakraborty:2024sco}, \cite{q}.  Using, the expressions for $b(r),~\Phi(r)$ and $f(Q)$ in the field equations (\ref{eq12})-(\ref{eq14}) we now check various ECs namely, NEC,WEC,SEC and DEC and show them graphically in FIG. (\ref{f1}) 

	Remark 1: In case of the first WH solution, although the energy density is positive but all the ECs fail to hold. Thus, matter supporting this WH formation is exotic.
	\subsection{Solution II}
\begin{figure}[h!]
		\centering		
		\begin{minipage}{0.3\textwidth}
			\includegraphics[height=5cm,width=6.5cm]{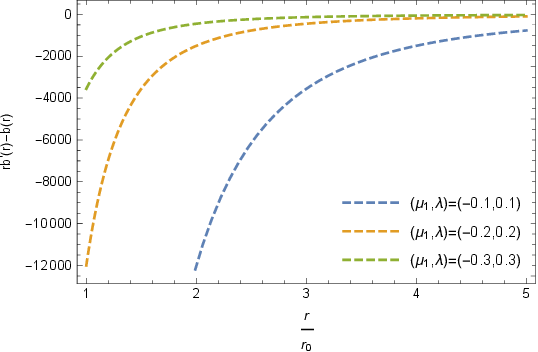}
		\end{minipage}
		\begin{minipage}{0.85\textwidth}\caption{Flairing out condition in case of Solution II for various choices of the model parameters specified in each panel.}\label{f2.1}
		\end{minipage}
	\end{figure}

\begin{figure}[h!]
		\begin{minipage}{0.3\textwidth}
			\centering\includegraphics[height=5cm,width=5.5cm]{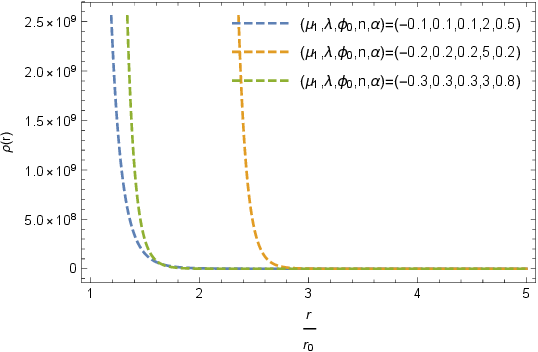}
		\end{minipage}~~~~~~~
		\begin{minipage}{0.3\textwidth}
			\centering\includegraphics[height=5cm,width=5.5cm]{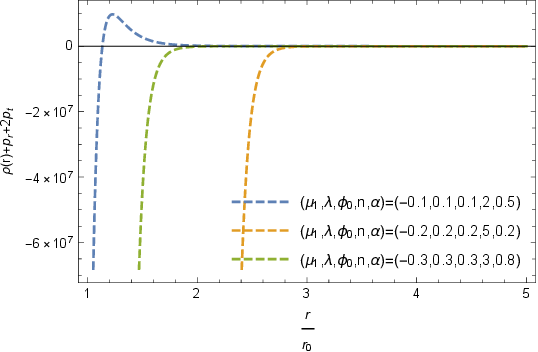}
		\end{minipage}\hfill
		\begin{minipage}{0.3\textwidth}
			\centering\includegraphics[height=5cm,width=5.5cm]{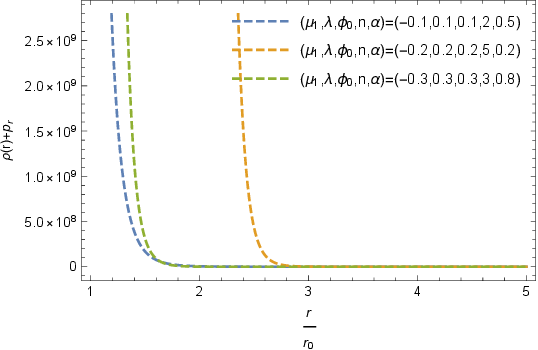}
		\end{minipage}
		\begin{minipage}{0.3\textwidth}
			\centering\includegraphics[height=5cm,width=5.5cm]{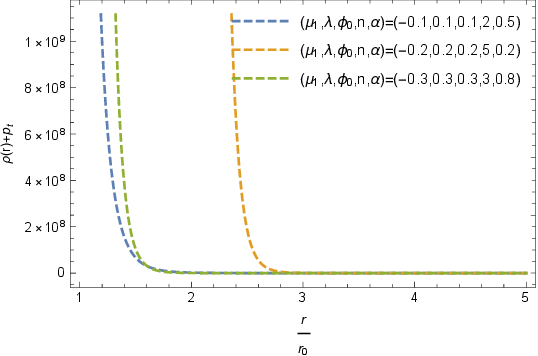}
		\end{minipage}
		\begin{minipage}{0.3\textwidth}
			\centering\includegraphics[height=5cm,width=5.5cm]{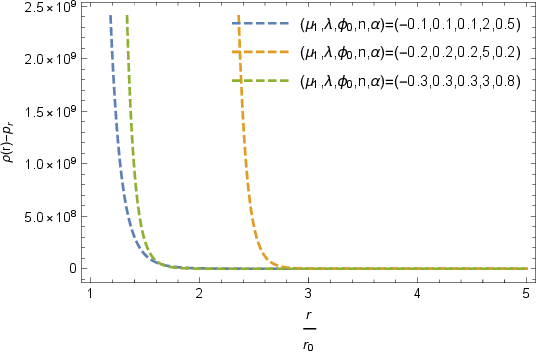}
		\end{minipage}
		\begin{minipage}{0.3\textwidth}
			\centering\includegraphics[height=5cm,width=5.5cm]{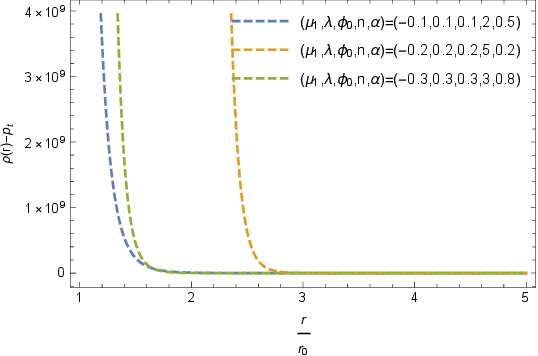}
		\end{minipage}
		\begin{minipage}{0.85\textwidth}\caption{NEC,WEC,SEC and DEC in case of Solution II for various choices of the model parameters specified in each panel.}\label{f2.2}
		\end{minipage}
	\end{figure}

	Now, we impose the constraint $2r\Phi'(r)+r^{2}\Phi''(r)+r^{2}\Phi'^{2}(r)=0$ to get the red-shift function as
	\begin{equation}
		\Phi(r)=\ln\left(1-\dfrac{\Phi_{0}}{r}\right)\label{eq21}
	\end{equation}
	Next, to determine $b(r)$, we consider the functional relation
	\begin{equation}
		\dfrac{\left(1-\dfrac{b(r)}{r}\right)}{r}-\dfrac{(b'r-b)}{4r^{2}}=\dfrac{1}{r}\left(\mu_{0}+\mu_{1}e^{-\lambda r}\right)
	\end{equation} where, $\mu_{0},~\mu_{1},~\lambda$ are constants. Solving the above differential equation we get,
	\begin{equation}
		b(r)=r(1-\mu_{0})-\dfrac{4\mu_{1}}{\lambda}e^{-\lambda r}\left(1+\dfrac{3}{\lambda r}+\dfrac{6}{\lambda^{2}r^{2}}+\dfrac{6}{\lambda^{3}r^{3}}\right)
	\end{equation}
	Feasible shape function and the asymptotic flatness imply $\mu_{0}=1,~\lambda\geq0,~\mu_{1}<0$ so that the solution becomes 
	\begin{equation}
		b(r)=-\dfrac{4\mu_{1}}{\lambda}e^{-\lambda r}\left(1+\dfrac{3}{\lambda r}+\dfrac{6}{\lambda^{2}r^{2}}+\dfrac{6}{\lambda^{3}r^{3}}\right)\label{eq24}
	\end{equation}
	Using equations (\ref{eq11}), (\ref{eq21}) and (\ref{eq24}) we get,
	\begin{equation}
		Q=-\dfrac{2e^{-\lambda r}(r+\Phi_{0})(e^{\lambda r}r^{4}\lambda^{4}+4\mu_{1}(6+6\lambda r+3\lambda^{2}r^{2}+\lambda^{3}r^{3}))}{r^{6}(r-\Phi_{0})\lambda^{4}}\label{eq25*}
	\end{equation}
	The flairing out condition and ECs in case of Solution II have been checked and shown graphically in FIG. (\ref{f2.1}) and (\ref{f2.2}) considering the same functional form of $f(Q)$ used in the previous case.

	Remark 2: In this WH solution, all ECs except the SEC hold. Thus, matter is exotic but not phantom, it is quintessence in nature.
	
	 Remark 3: It is to be noted that, section 2.2 deals with the ECs in GR involving $\rho,~p_{r},~p_{t}$. However, using the field equations (\ref{eq12}), (\ref{eq13}) and (\ref{eq14}) we have found the expressions for modified $\rho,~p_{r},~p_{t}$ which contains $Q,~f(Q),~f_{Q}$ etc. The expressions for $Q=Q(r)$ for both the WH solutions are given by equations (\ref{eq20*}) and (\ref{eq25*}). Accordingly, the ECs have been plotted w.r.t $\dfrac{r}{r_{0}}$ for various choices of the model parameters that implicitly carry the effect of non-metricity. The key take-away from the ECs analysis is that in case of WH1, negative non-metricity (indicative of repulsive gravity) plays the role of exotic matter while the same negative non-metricity plays the role of quintessence fluid (although exotic, but not phantom) in WH2.
	\section{Shadows of $f(Q)$ WHs: Analytic Prescription}
	\subsection{WH configuration and null geodesics}
	The null geodesic equation in the vicinity of WH throat is given by 	\cite{egb}
	\begin{equation}
		\dfrac{dK^{\alpha}}{d\Lambda}+\Gamma_{\mu\nu}^{\alpha}K^{\mu}K^{\nu}=0
	\end{equation}
	where, $K^{\alpha}$ is a null vector i.e, $K^{\alpha}=\dfrac{dx^{\alpha}}{d\Lambda}$ is tangent to the null geodesic satisfying $K^{\alpha}K_{\alpha}=0$, $\Lambda$ is the affine parameter along the geodesics. In Hamilton-Jacobi approach, the geodesic equations can be written as
	\begin{equation}
		\dfrac{dS}{d\Lambda}=-\dfrac{1}{2}g^{ab}\dfrac{\partial S}{\partial x^{a}}\dfrac{\partial S}{\partial x^{b}}
	\end{equation} where $S$ is called the Jacobi function. As, the WH space-time is static spherically symmetric so the geodesic motion is associated with two integrals of motion namely Energy ($E$) of the particle and angular momentum ($L$) about the axis of symmetry. We assume another conserved quantity namely, the Carter constant/ separation constant denoted by $\lambda_{0}$. Thus, we have
	\begin{equation}
		S=\dfrac{1}{2}m_{0}^{2}\Lambda-Et+L\phi+R(r)+T(\theta)
	\end{equation} with $m_{0}$= mass of the test particle ($m_{0}=0$ for null particle), $t$ is the usual time-like coordinate, $\phi$ is space-like coordinate that identifies the orbits of the space-like killing field. $R$ and $T$ are the functions of the variables in the argument only. Using the WH space-time metric, the Hamilton-Jacobi (HJ) equation explicitly takes the form
	\begin{equation}
		-E^{2}e^{-2\Phi}+\left(1-\dfrac{b(r)}{r}\right)\left(\dfrac{dR}{dr}\right)^{2}+\frac{1}{r^{2}}\left(\dfrac{dT}{d\theta}\right)^{2}+\dfrac{L^{2}}{r^{2}\sin^{2}\theta}=0
	\end{equation}
	Carter constant $\lambda_{0}$ is used to get,
	\begin{equation}
		\left(\dfrac{dR}{d\Lambda}\right)^{2}=\dfrac{E^{2}e^{-2\Phi}-\dfrac{\lambda_{0}}{r^{2}}}{\left(1-\dfrac{b(r)}{r}\right)}=A(r)
	\end{equation} and,
	\begin{equation}
		\left(\dfrac{dT}{d\theta}\right)^{2}=\left(\lambda_{0}-\frac{L^{2}}{\sin^{2}\theta}\right)=B(\theta)
	\end{equation}
	The HJ formalism along with the expressions for momentum from the Lagrangian formulation one has
	\begin{eqnarray}
		\dfrac{dt}{d\Lambda}=Ee^{-2\Phi}\label{eq32}\\
		\dfrac{d\phi}{d\Lambda}=\dfrac{L}{r^{2}\sin^{2}\theta}\label{eq33}
	\end{eqnarray}
	Now, 
	\begin{equation}
		\dfrac{1}{\left(1-\dfrac{b(r)}{r}\right)}\dfrac{dr}{d\Lambda}=p_{r}=\dfrac{dR}{dr}=\sqrt{A(r)}
	\end{equation}
	\begin{equation}
		\dfrac{dr}{d\Lambda}=\sqrt{\left(1-\dfrac{b(r)}{r}\right)\left(E^{2}e^{-2\Phi}-\dfrac{\lambda_{0}}{r^{2}}\right)}=A_r(r)\label{eq35}
	\end{equation} and,
	\begin{equation}
		\dfrac{d\theta}{d\Lambda}=\dfrac{1}{r^{2}}\sqrt{\lambda_{0}-\dfrac{L^{2}}{\sin^{2}\theta}}=B_{\theta}(\theta)\label{eq36}
	\end{equation}
	For classical solutions, the functions $A_{r}(r)$ and $B_{\theta}(\theta)$ should be real and the null geodesics are characterized by the conserved quantities $E,~L$ and $\lambda_{0}$ out of which only two are independent. Thus, one may reasonably define $\mu=\dfrac{L}{E}$ and $\nu=\dfrac{\lambda_{0}}{E^{2}}$.
	\subsection{What is a WH shadow?}
	Let us suppose that among the distant space-time regions those are connected by $f(Q)$ WHs, one space-time region is illuminated by a source of light and there is no light source near the throat of the WH in the other region. In this situation, there are two possibilities
	\begin{enumerate}
		\item Photons enter the WH and pass through its throat.
		\item Photons scatter away from the WH to infinity.
	\end{enumerate}
	Thus, an observer at infinity in the first region will only be able to see the scattered ones while the photons captured by the WH will appear as dark spot in the luminous background. This is termed as \textit{Shadows} of the WH. From mathematical point of view, in order to have scattered photons the radial motion should have a turning point characterized by 
	\begin{equation}
		\frac{dr}{d\Lambda}=0,~V_{eff}=0
	\end{equation}
	Further, the critical orbit between the scattered and plunged orbits is characterized by the maximum of the effective potential. This critical orbit is spherical and it is unstable in nature as due to a small perturbation / distortion it may either escape or be captured orbits. Therefore, $V_{eff}=0=\dfrac{dV_{eff}}{dr}$ and $\dfrac{d^{2}V_{eff}}{dr^{2}}\leq0$ characterize the critical orbit. So, for the present $f(Q)$ WH configuration, $r_{ph}$, the radius of the photon sphere locates the apparent size of the photon ring and satisfies 
	\begin{eqnarray}
		r\Phi'(r)=1\label{eq38}\\
		\implies e^{\Phi(r)}=\Phi_{0}r
	\end{eqnarray}
	$r_{ph}$ is the largest positive real root of the equation $r\Phi'(r)=1$ and the impact parameter $\nu$ satisfies the shadow radius as $r_{sh}=\sqrt{\nu}=re^{-\Phi(r)}|_{r=r_{ph}}$, where $r_{sh}$ is the radius of the shadow. Conveniently, we assume that the observer is located away from the WH so that the WH shadow radius can be expressed as \cite{egb}
	\begin{equation}
		r_{sh}=r_{ph}e^{-\Phi(r_{ph})}=\sqrt{X^{2}+Y^{2}}
	\end{equation} where $(X,Y)$ is the celestial coordinates in the observer's frame.
	
	Let $r_{0}$ be the distance of the observer from the WH and $\theta_{0}$ be the angle of inclination (angular coordinate of the observer). Thus, $(r_{0},\theta_{0})$ is the location of the observer. Then, $(X,Y)$ identifies the boundary curves of the WH shadow (the apparent shape) and related to $(r_{0},\theta_{0})$ by
	\begin{eqnarray}
		X=\lim_{r_{0}\rightarrow\infty}\left(-r_{0}^{2}\sin\theta_{0}\right)\dfrac{d\phi}{dr}\\
		Y=\lim_{r_{0}\rightarrow\infty}\left(r_{0}^{2}\dfrac{d\theta}{dr}\right)
	\end{eqnarray} i.e,
	\begin{eqnarray}
		X=-\dfrac{\mu}{\sin\theta_{0}}\\
		Y=\sqrt{\nu-\dfrac{\mu^{2}}{\sin^{2}\theta_{0}}}
	\end{eqnarray}
	In doing these deductions, expressions for four velocity given by equations (\ref{eq32})-(\ref{eq36}) have been used and $\lim_{r_{0}\rightarrow\infty}e^{-2\phi(r_{0})}$= constant has been considered. Using the above interrelations among the celestial coordinates and the impact parameter one may form the shadow of the $f(Q)$ WH. To represent diagramatically, one has to plot $X~vs ~Y$ to identify the boundary of the shadow and in doing so we consider the equatorial plane $\theta_{0}=\dfrac{\pi}{2}$. 
	\subsection{Shadow of WH1 and WH2 in the backdrop of $f(Q)$ gravity: Effect of gravity via non-metricity}
	While black hole and wormhole shadows have distinct structural differences, yet both are fascinating concepts that may present unique observational features. The primary difference between them lies in the space-time geometry they represent. A black hole shadow is formed due to the gravitational bending of light around the event horizon, the boundary beyond which nothing,  not even light, can escape. The size, shape, event horizon, and photon ring primarily define a typical black hole shadow, such as those of supermassive black holes like $M87*$ and $SgrA*$. Observations from the Event Horizon Telescope (EHT) reveal that their shadows appear mostly circular, with slight perturbations imposed by spin, photon ring structure, and surrounding plasma interactions \cite{Afrin:2021wlj}, \cite{Abdujabbarov:2016hnw}, \cite{Afrin:2021imp}, \cite{Kumar:2020yem}.
	
	On the contrary, Wormholes do not have a conventional event horizon. Instead, they feature a structure that permits light to pass through from one side to the other. As a result, their shadows may not display the well-defined boundary seen in black holes but could instead show more intricate patterns based on the wormhole's geometry. If a wormhole is traversable, light can potentially escape from the opposite side, leading to complex emission patterns observable in high-resolution images or through time-dependent variations in brightness as matter moves around or through the wormhole.
	
	Distinguishing black hole and wormhole shadows is challenging, particularly with current observational technology. The EHT, which captured the first images of $M87*$ and $SgrA*$, has limited resolution \cite{Afrin:2021wlj}. While black hole shadows have been detected, there is currently no definitive way to include the possibility of wormholes based on observational data alone. Typically, black hole shadows appear smooth and circular, with subtle distortions due to spin and a surrounding photon ring.
	
	Building on the EHT’s findings on black hole shadows, this study explores the shadows of viable wormhole solutions within the framework of $f(Q)$ gravity. We examine the shadows in both the WH solutions so obtained by constraining  non-metricity so that wormhole shadows in $f(Q)$ gravity resemble the observed shadows of $M87*$ and $SgrA*$. Differentiating wormholes and black holes would require extremely high-resolution observations and precise modeling of gravitational lensing, as the features of their shadows may overlap within the limits of current observational capabilities.
	
\begin{figure}[h!]
		\centering	\begin{minipage}{0.3\textwidth}
			\includegraphics{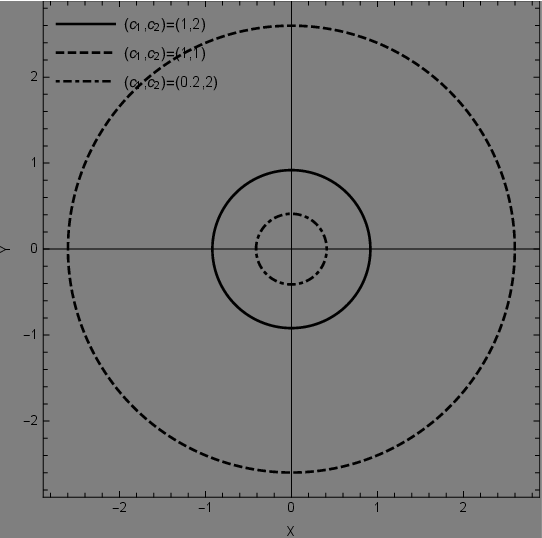}
		\end{minipage}
		\caption{Shadows for various model parameters in case of WH1. Variation of model parameters have been shown by three distinct lines: dotted, dashed and dot-dashed.}\label{f3.1}
	\end{figure}

	In the present context, corresponding to Solution I and II, we represent the effective potential i.e, $V_{eff}$ and show the variation of the WH shadows with non-metricity that carries the effect of gravity in the present gravity theory in FIG . (\ref{f3.1}) and (\ref{f3.2}).
	\begin{figure}[h!]
		\begin{minipage}{0.3\textwidth}
			\includegraphics[height=4.5cm,width=7cm]{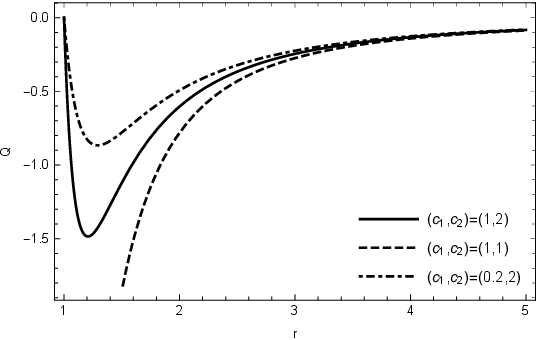}
		\end{minipage}\hfill
		\begin{minipage}{0.3\textwidth}
			\includegraphics[height=4.5cm,width=7cm]{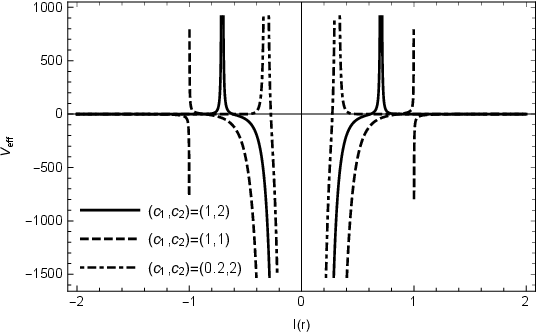}
		\end{minipage}
		\caption{Variation of non-metricity $Q$ with $r$ and $V_{eff}$ with proper radial distance $l(r)=\pm r$ for WH1.}
	\end{figure}
	\begin{figure}[h!]
		\centering	\begin{minipage}{0.3\textwidth}
			\includegraphics{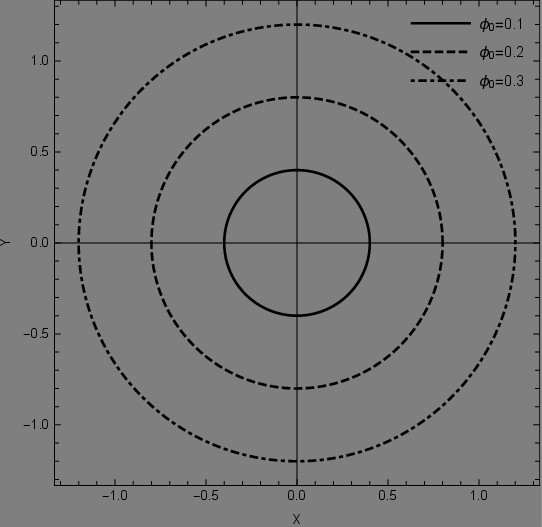}
		\end{minipage}
		\caption{Shadows for various model parameters in case of WH2. Variation of model parameters have been shown by three distinct lines: dotted, dashed and dot-dashed.}\label{f3.2}
	\end{figure}
	\begin{figure}[h!]
		\begin{minipage}{0.3\textwidth}
			\includegraphics[height=4.5cm,width=7cm]{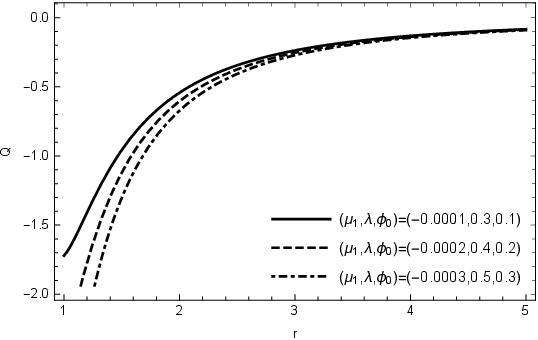}
		\end{minipage}\hfill
		\begin{minipage}{0.3\textwidth}
			\includegraphics[height=4.5cm,width=7cm]{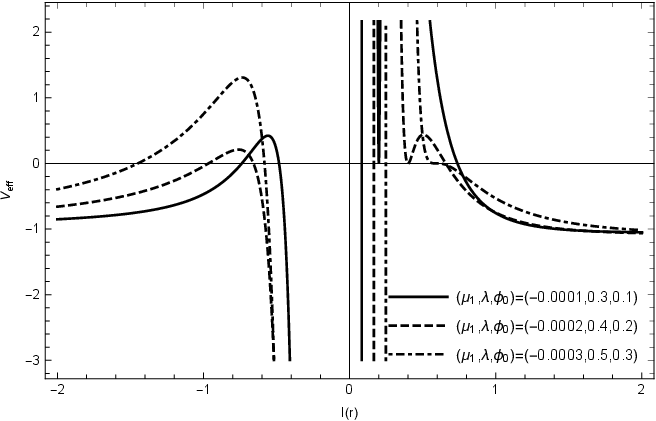}
		\end{minipage}
		\caption{Variation of non-metricity $Q$ with $r$ and $V_{eff}$ with proper radial distance $l(r)=\pm r$ for WH2.}
	\end{figure}

	Remark: In both the WHs, negative non-metricity allows shadow formation. Further, the shadow becomes larger in size (radius of the shadow increases) as non-metricity becomes more negative.  Non-metricity certainly affects geometric deformation of light paths. It allows how geodesics are defined . Photons do not necessarily follow the geodesics of the levi-civita connection. $Q<0$ can enhance the gravitational lensing effects, leading to larger photon deflection which helps in the formation of shadow. Further, if we look at the modified field equations in $f(Q)$ gravity as $G_{\mu\nu}=T_{\mu\nu}^{eff}=T_{\mu\nu}^{matter}+T_{\mu\nu}^{Q}$, then the non-metricity induced part can be interpreted effective energy momentum contribution. So when, $Q<0$ it often leads to repulsive contributions or negative pressure which are crucial in supporting feasible WH throat and light trapping regions.
		A shadow is formed if there is a photon sphere or some regions that trap photons. $Q<0$ can distort the $V_{eff}$ for photons creating a stable circular orbit -- a shadow.
	
 Thus, $Q<0$ has the following physical implications :
		\begin{enumerate}
			\item $Q<0$ deforms space-time geometry significantly enough to trap photons.
			\item $Q<0$ helps to satisfy flairing-out conditions.
			\item $Q<0$, in case of $f(Q)$ WHs generates photon rings/ shadows when solving null geodesics.
			\item From a physical perspective, $Q<0$ adds repulsive gravity effects and instability of null geodesics needed for photon rings and subsequent formation of shadows.
	\end{enumerate}
	\section{Quasinormal modes and Grey body factors of traversable WHs: Bridging theory and observations}
\begin{figure}[h!]
		\centering	\includegraphics{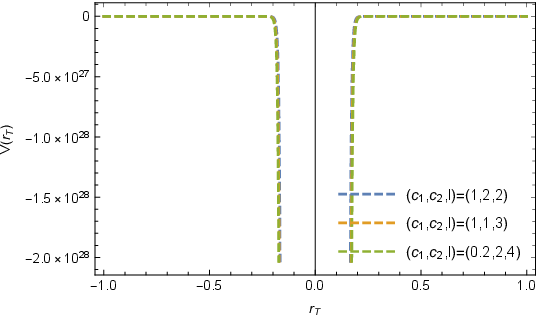}
		\caption{$V(r_{T})$ vs $r_{T}$ for various choices of the parameters specified in panel}\label{f4}
	\end{figure}

	In the context of gravitational waves (GW) astronomy, the study of quasinormal modes (QNMs) and grey body factors for WHs has enhanced interest in recent times \cite{qnms}, \cite{gb}. These quantities identify the response of the space-time to perturbations and its observational characteristics. The damped oscillations of space-time are represented by QNMs and are uniquely characterized by the geometry of the WH and boundary conditions. On the other hand, grey body factors measure the transmission probability of waves propagating through the WH's effective potential barrier. Thus, it is speculated that both the QNMs and grey body factors may be considered as tools to distinguish WHs from BHs in the context of GW astronomy and related observational studies. Further, grey body factors may exhibit greater stability against deformations of the space-time geometry near the horizon while the overtones of QNMs are highly sensitive to such deformations. One may perturb a WH geometry given by (\ref{eq9}), considering small deviations in the background space-time geometry or equivalently by examining the evolution of test fields in the underlying geometry. As a result, the perturbation field follows Schrödinger-like  equation as \cite{qnms}, \cite{gb}
	\begin{equation}
		\left(\dfrac{d^{2}}{dr_{T}^{2}}+\omega^{2}-V(r_{T})\right)\Psi(r_{T})=0\label{eq45}
	\end{equation}
	Here, $\omega$ stands for the frequency of oscillation, $V$ is the effective potential that depends on the WH geometry and $r_{T}$ is the tortoise coordinate having expression 
	\begin{equation}
		r_{T}(r)=\int_{r_{0}}^{r}\dfrac{dr'}{e^{\Phi}\sqrt{1-\dfrac{b(r')}{r'}}}
	\end{equation}  
	The tortoise coordinate $r_{T}$ identifies the throat at $r_{T}=r_{0}$ while $r_{T}=\pm \infty$ corresponds to the asymptotically flat regions far from the throat. The explicit form of the effective potential is given by 
	\begin{equation}
		V(r)=e^{2\Phi}\left(\dfrac{l(l+1)}{r^{2}}-\dfrac{(rb'-b)}{2r^{3}}+\dfrac{\Phi'(r)}{r}\left(1-\dfrac{b(r)}{r}\right)\right)\label{eq47}
	\end{equation} for massless scalar field and 
	\begin{equation}
		V(r)=e^{2\Phi}\dfrac{l(l+1)}{r^{2}}
	\end{equation} for electromagnetic field. The variation of the potential for electromagnetic wave with tortoise coordinate for WH1 is given in FIG. (\ref{f4}). The plot shows that, potential has maximum value at the throat radius $r_{0}=1$. Thus, we can get the quasinormal modes and grey body factors associated with WH1 using the methodology discussed later in this section. However, it is not feasible to find the expression for the potential as function of tortoise coordinate in case of WH2 due to the complex nature of the shape function leading to mathematical complications.

	The above effective potential as a function of the tortoise coordinate framework represents a positive definite potential barriers having a maximum at the throat. Here, $l$ (the angular momentum quantum no.) represents the multipole number. 
	
	The QNMs of WHs are the solutions of the above wave equation (\ref{eq45}) with the purely outgoing boundary conditions at infinity i.e, 
	\begin{equation}
		\Psi(r_{T})\approx \Psi_{I} e^{\pm i\omega r_{T}}, ~r_{T}\rightarrow\mp\infty\label{eq49}
	\end{equation}
	The above equation indicates purely outgoing waves at both asymptotic regions without any incoming wave from either side. This solution is termed as QNM having $\omega=Re(\omega)+i Im(\omega)$. It is to be noted that, the QNM gives information about the evolution of the perturbed field as follows: the real part of $\omega$ identifies the oscillation of the signal while the imaginary part measures the damping factor for the energy loss of gravitational radiation. More specifically, \\
	$Im(\omega)>0$, unstable perturbations with exponential growth of the perturbed field\\
	$Im(\omega)\leq 0$, implies stability of the perturbed field.
	This is because, from equation (\ref{eq49}) and using $\omega=Re(\omega)+i Im(\omega)$ we have 
		\begin{equation}
			\Psi(r_{T})\approx e^{\pm i Re(\omega)r_{T}}\times e^{\mp Im(\omega)r_{T}}
		\end{equation}
		Now, to find a finite value of $|\Psi|^{2}$ (the probability measure) we need the absolute value of  $e^{\pm i Re(\omega)r_{T}}$ and $e^{\mp Im(\omega)r_{T}}$, the former is unity and the later is  bounded as $r_{T}\rightarrow \mp \infty$ if and only if $Im(\omega)\leq 0$.
	\\
	On the other hand, due to Schrödinger like nature of wave equation (\ref{eq45}) the WKB approximation is preferable in this arena. Accordingly, the formula for the QNM frequencies to the $k-th$ order in perturbation takes the form 
	\begin{equation}
		i\dfrac{\omega^{2}-V_{0}}{\sqrt{-2V_{0}''}}-\sum_{i=2}^{K}\Lambda_{i}=k+\dfrac{1}{2}
	\end{equation}
	Here, $V_{0}$ is the maximum of the potential , its second order derivative w.r.t the tortoise coordinate is denoted by $V_{0}$, all higher order corrections are contained in $\Lambda$ and $k$ stands for the order of the WKB approximation. It is important to note that larger $k$ does not give a better approximation for the quasinormal frequencies. In fact, the value of $k$ for best value of quasinormal frequency is not unique it depends on $(n,l)$.
	
	On the other hand, to examine the wave scattering processes in the vicinity of the WH, one has to take into account the interaction of waves with potential barrier surrounding the WH (compact object). Consequently, waves are partially reflected off the barrier and partially transmitted through it. The grey body factors take care of both these events and it does not depend on whether the wave originates in the asymptotic region in the other universe or it arrives from spatial infinity. Thus, here the boundary conditions are modified as
	\begin{eqnarray}
		\Psi=e^{-i\Omega_{r_{T}}}+R e^{i\Omega_{r_{T}}},~r_{T}\rightarrow +\infty\nonumber\\
		\Psi=T e^{-i\Omega_{r_{T}}},~r_{T}\rightarrow-\infty
	\end{eqnarray} with $R$ and $T$ as reflection and transmission coefficients. Here, the frequency $\Omega$ is real and continuous for scattering phenomena and is distinct from the complex and discrete quasi normal mode frequency. The transmission coefficient $T$ is usually termed as the grey body factor as it measures the fraction of the wave which traverses the potential barrier and as a result it contributes to the emission of radiation from the compact objects (BH or WH). Thus, one may write the grey body factor corresponding to the angular momentum number $l$ as
	\begin{equation}
		\Gamma_{l}(\Omega)=|T|^{2}=1-|R|^{2}
	\end{equation}
	Thus, using WKB expression for the grey body factor one has
	\begin{equation}
		\Gamma_{l}(\Omega)=\dfrac{1}{1+e^{2\pi i k}}
	\end{equation}
	Now, the effective potential can be expanded in powers of $l$
	\begin{equation}
		V(r_{T})=l^{2}V_{0}(r_{T})+lV_{1}(r_{T})+l^{-1}V_{2}(r_{T})+...\label{eq54}
	\end{equation}
	and as a result the eikonal approximation can be derived from the first order WKB approximation as
	\begin{equation}
		\Omega=l\sqrt{V_{00}}-ik\sqrt{\dfrac{-V_{00}''}{2V_{00}}}+O(l^{-1})
	\end{equation}
	Thus, quasinormal modes (with $n=0,~k=\dfrac{1}{2}$) can be estimated as (with $n=0,~k=\dfrac{1}{2}$) 
	\begin{equation}
		\Omega_{0}=l\sqrt{V_{00}}-\dfrac{i}{2}\sqrt{\dfrac{-V_{02}}{2V_{00}}}+O(l^{-1})
	\end{equation}
	\begin{equation}
		Re(\omega_{0})=l\sqrt{V_{00}}+O(l^{-1})\label{eq57}
	\end{equation}and,
	\begin{equation}
		Im(\omega_{0})=-\dfrac{1}{2}\sqrt{\dfrac{-V_{02}}{2V_{00}}}+O(l^{-1})\label{eq58}
	\end{equation}
	As a consequence, $k$ can be expressed as a function of the real frequency $\Omega$
	\begin{equation}
		-ik=\dfrac{\Omega^{2}-l^{2}V_{00}}{l\sqrt{-2V_{00}}}+O(l^{-1})=-\dfrac{\Omega^{2}-Re(\omega_{0})^{2}}{4 Re(\omega_{0})Im(\omega_{0})}+O(l^{-1})
	\end{equation}
	Hence, one may associate the transmission coefficient to the grey body factors $\Gamma_{l}(\Omega)$ with the fundamental mode $\omega_{0}$ as 
	\begin{equation}
		\Gamma_{l}(\Omega)=|T|^{2}=\left(1+e^{^{2\pi\frac{\Omega^{2}-Re^{2}\omega_{0}}{4Re(\omega_{0})Im(\omega_{0})}}}\right)\label{eq61}
	\end{equation}
	It is to be noted that the above relation is exact in the eikonal limit $l\rightarrow \pm \infty$ and an approximate one for small $`l'$. 
	\section{Quasinormal modes and Grey body factors in $f(Q)$ Wormholes}
		QNMs are the characteristic oscillations of space-time in response to perturbations. The real part gives the oscillation frequency and the imaginary part gives the damping rate. QNMs are crucial to study in the context of TWHs in $f(Q)$ gravity theory. This is because, QNMs enable us to test whether small perturbations decay (leading to stable WH) or grow (unstable WH). Since, $f(Q)$ gravity modifies the dynamics of perturbations (via the non-metricity), its effect can be understood by seeing the deviation in QNM spectra from that predicted in GR. If a WH does exist, then gravity waves produced by its perturbations would carry QNM imprints (just like finger prints, they encode information about the mass, size of the throat, radius of the WH shadow, red-shift and the form of $Q$). In order to potentially distinguish WHs from BHs, LIGO and VIRGO detectors (high precision instruments) are required \cite{Isi:2021iql}.
		
		On the other hand, grey body factors describe how a wave interacts with the WH's effective potential barrier. The wave is partially transmitted and partially reflected. Actually, grey body factors quantify the transmission probability. Grey body factors exhibit how much energy escapes through the WH versus how much is being reflected. Since, WHs are without horizon, so they can radiate in both the directions unlike BHs. QNMs track natural resonances while grey body factors measure scattering behavior. This is the distinction, although they are theoretically interconnected. Grey body factors influence the thermal emission spectra. For BHs, this is Hawking radiation. Studying QNMs and grey body factors help constrain the $f(Q)$ model to observations. For WHs, grey body factors impact the hypothetical emission from accretion or Hawking-like processes. Like QNMs, grey body factors also hint about how QNMs, GBFs also give hint about how ``transparent" or ``resonant" the geometry is towards perturbations.
		
\begin{figure}[h!]
			\begin{minipage}{0.3\textwidth}
				\includegraphics[height=5.5cm,width=7.5cm]{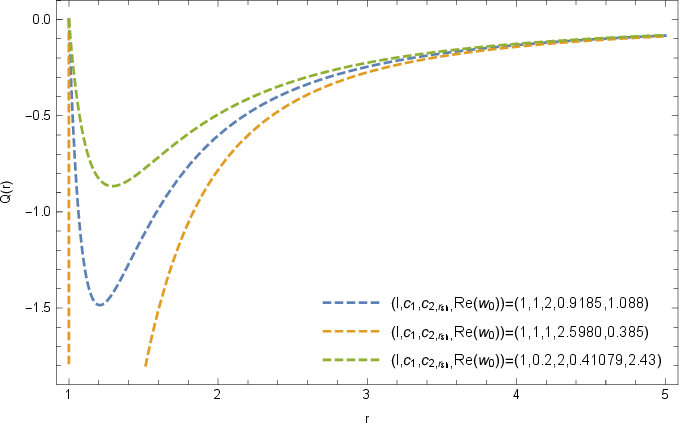}
			\end{minipage}~~~~~~~~~~~~~~~~~~~~~~~~~~~~~~~~~~~~~~~~~~
			\begin{minipage}{0.3\textwidth}
				\includegraphics[height=5.5cm,width=7.5cm]{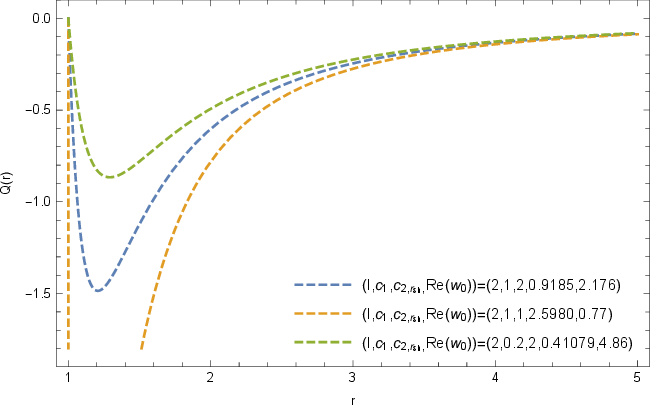}
			\end{minipage}\\ \\
			\centering \begin{minipage}{0.3\textwidth}
				\includegraphics[height=6.5cm,width=8.5cm]{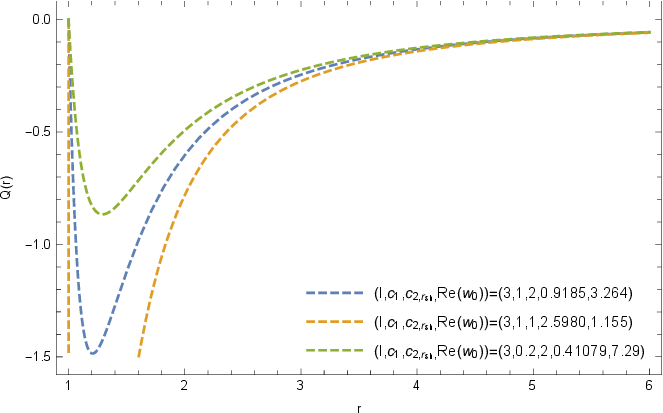}
			\end{minipage}
			\caption{QNM spectra for WH1 showing effect of non-metricity and radius of shadow}\label{F9}
		\end{figure}

\begin{figure}[h!]
			\begin{minipage}{0.3\textwidth}
				\includegraphics[height=5.5cm,width=7.5cm]{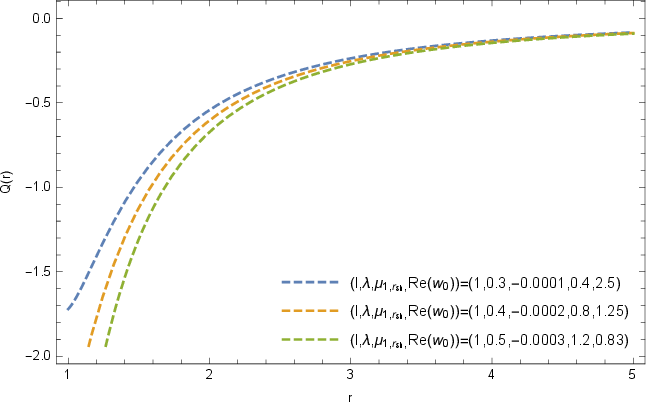}
			\end{minipage}~~~~~~~~~~~~~~~~~~~~~~~~~~~~~~~~~~~~~~~~~~
			\begin{minipage}{0.3\textwidth}
				\includegraphics[height=5.5cm,width=7.5cm]{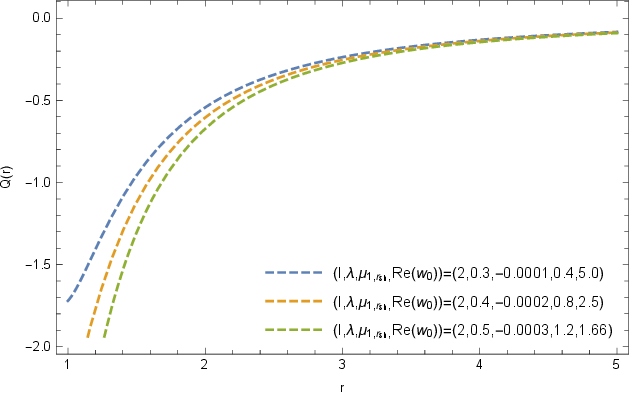}
			\end{minipage}\\ \\
			\centering \begin{minipage}{0.3\textwidth}
				\includegraphics[height=6.5cm,width=8.5cm]{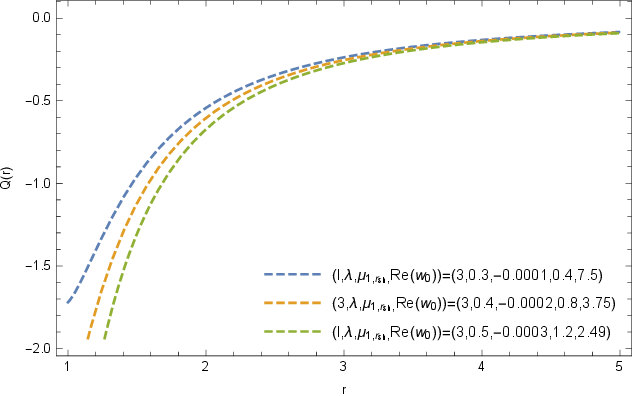}
			\end{minipage}
			\caption{QNM spectra for WH2 showing effect of non-metricity and radius of shadow}\label{F10}
		\end{figure}

		Comparing equations (\ref{eq47}) and (\ref{eq54}), we have
		\begin{equation}
			V_{0}(r_{T})=\dfrac{e^{2\Phi(r_{T})}}{r_{T}^{2}}
		\end{equation}
		The maximum value of $V_{0}(r_{T})$ occurs at the radius of the photon ring $r_{ph}$ (this is because, $V_{0}'(r_{T})=0\implies r_{T}=r_{ph}$ where equation (\ref{eq38}) has been used). So, $V_{00}=V_{0}(r_{ph})$ and $V_{02}=V_{00}''=0$. Thus, using equations (\ref{eq57}) and (\ref{eq58}), the expressions for $Re(\omega_{0})$ and $Im(\omega_{0})$ are given by 
		\begin{eqnarray}
			Re(\omega_0)=l\dfrac{e^{\Phi(r_{ph})}}{r_{ph}}\\
			Im(\omega_{0})=0
		\end{eqnarray}
		For WH1, the QNM spectra has the expression
		\begin{equation}
			Re(\omega_{0})=l\left(\dfrac{c_{2}-\dfrac{c_{1}}{r_{ph}^{2}}}{r_{ph}}\right), ~~Im(\omega_{0})=0
		\end{equation}
		Substituting $c_{2}=\dfrac{Re(\omega_{0})\times r_{ph}}{l}+\dfrac{c_{1}}{r_{ph}^{2}}$ in equation (\ref{eq20*}) we analyze the effect of non-metricity in the QNM spectra for $l=1,2,3$ ($Re(\omega_{0})=0$ when $l=0$) and with radius of shadow given by ($r_{sh}=r_{ph}e^{-\Phi(r_{ph})}$) in FIG.(\ref{F9})
		Similarly, we calculate the QNMs for WH2 as follows 
		\begin{eqnarray}
			Re(\omega_{0})=l\left(\dfrac{1-\dfrac{\Phi_{0}}{r_{ph}}}{r_{ph}}\right)\\
			Im(\omega_{0})=0
		\end{eqnarray}
		so that substituting $\Phi_{0}=r_{ph}\times \left(1-\dfrac{r_{ph}Re(\omega_{0})}{l}\right)$ in equation (\ref{eq25*}) and using $r_{sh}=r_{ph}e^{-\Phi(r_{ph})}$ we can see the effect of non-metricity in QNM spectra for WH2 with radius of shadow as shown in FIG (\ref{F10}).

		Remark: In both WHs, we see that frequency of the oscillation (represented by $Re(\omega_{0})$ decreases as $Q$ becomes more negative or gravity is more repulsive. Further, there is a nice interrelation among non-metricity scalar, radius of the shadow and the frequency of oscillation as depicted in the plots. Therefore, more repulsive is the gravity, larger is the radius of the shadow and smaller will be the frequency of oscillation. Further, since $Im(\omega_{0})=0$ there is no damping effect and as $Im(\omega_{0}\leq 0)$ implies stability therefore the model under consideration is stable.
		Further, from the equation (\ref{eq61}) since we have $Im(\omega_{0})=0$ so $\Gamma_{l}(\Omega)\rightarrow 1$ only if $\Omega<\Omega_{0}$. That means, in order to have a finite (close to unity) value of grey body factor, the QNM frequencies should fall below the QNM frequency $\Omega_{0}$. If there exists a frequency that falls above $\Omega_{0}$, then $\Gamma_{l}$ corresponding to that frequency will be infinite (indicating infinite transmission). Thus, the QNM frequency $\Omega_{0}=Re(\omega_{0})+i Im(\omega_{0})$ acts as a critical frequency/ threshold frequency which in the present model reduces to $Re(\omega_{0})$ only as there is no effect of damping via $Im(\omega_{0})$.
	\section{Concluding Remarks}
	In this work, we have investigated traversable wormhole (TWH) solutions within the framework of $f(Q)$ gravity, emphasizing their fundamental characteristics, including traversability conditions, energy constraints, shadow formations, and quasinormal modes (QNMs). Our study highlights how the effect of non-metricity in $f(Q)$ gravity plays a crucial role in shaping the WH solutions and influencing their observational signatures.
	
	By employing a power-law model of 
	$f(Q)$, we derived WH solutions that satisfy necessary physical conditions such as asymptotic flatness and the flaring-out criterion. The energy conditions were analyzed in greater detail, revealing the nature of the matter supporting these WHs. While some solutions required exotic matter violating the null energy condition (NEC), others satisfied weaker energy constraints, hinting at the possibility of WH formation without extreme violations. The WH1 solutions are asymptotically flat and flairing-out conditions hold. Also, it is to be noted that the expression for $Q$ in case of WH1 is singular as $r\rightarrow \sqrt{\frac{c_{1}}{c_{2}}}$. Keeping this in mind, choices of $c_{1},~c_{2}$ are made. Further, the expression of $Q$ in the asymptotic limit i.e, $r\rightarrow\infty$ can lead to strong coupling especially in the perturbative regime. This is because, $Q\rightarrow 0$ leads to divergence of power law model $f(Q)=\alpha Q^{n}$ for $n<1$. Thus, we have done the full analysis choosing $n\geq 1$ to avoid this pathology. While, in the case of WH2 solution the expression for $Q$ shows exponential decay via $e^{\lambda r}$ (see FIG. 7 left). This leads to comparatively stronger asymptotic behavior. Physically, this form of $Q$ allows better control over pathologies. Unlike, the expression for $Q$ in case of WH1 (see FIG. 5 left) in case of WH2 $Q$ decays faster and parameters like $\lambda,~\mu_{1}<0$ help regulate the shape and ECs. The ECs for WH2 are better behaved where all except the SEC hold good indicating a quintessence-like matter rather than a phantom matter.
	
	Furthermore, we examined the properties of these WH shadows, shedding light on how non-metricity affects their observational appearance. Unlike the black hole (BH) shadows, WH shadows exhibit distinct features that could be used to differentiate them in astrophysical observations. Our results suggest that negative non-metricity enhances the shadow size, providing an interesting avenue for testing alternative gravity models against real astronomical data, such as the images captured by the Event Horizon Telescope (EHT). In the present context of $f(Q)$ WHs, we see that shadow size grows larger as non-metricity becomes more negative (gravity is more repulsive). This is a direct evidence of space-time modification in the absence of metricity.

	Authors in \cite{Gogoi:2023fow} have explored QNMs and grey body factors in symmergent black hole and found that the  quasinormal model spectrum and grey body factor are sensitive to the symmergent parameter. On the other hand, in \cite{Konoplya:2023ppx} the authors studied QNMs and grey body factors of regular black holes with a scalar hair from the Effective Field Theory and derived the analytical formula for quasinormal modes in the eikonal regime. In addition, they calculated grey-body factors and showed that the regular Hayward black hole with a scalar hair has a smaller grey-body factor than the Schwarzschild one. In \cite{Guo:2023nkd}, three methods have been adopted to solve the quasinormal frequencies in the frequency domain and greybody factor has been calculated by Wentzel-Kramers-Brillouin (WKB) method. The effect of the Lorentz-violating parameter on the quasinormal modes and greybody factor has also been studied. Moreover, in \cite{Singh:2024nvx} the authors have studied the scalar and Dirac fields perturbation of Schwarzschild-de Sitter-like black hole in bumblebee gravity model. The effective potentials, greybody factors and quasinormal modes of the black hole have been investigated by using the Klein–Gordon equation and Dirac equation. Most importantly, The impact of Lorentz invariance violation parameter $L$ and cosmological constant $\Lambda$ to the effective potential, greybody factors and quasinormal modes have been analyzed for different modes. The authors in \cite{Kanzi:2021cbg} have investigated the QNMs and the greybody factors of the Kerr-like black hole spacetime obtained from the bumblebee gravity model and and discussed  impacts of the LSB on the bosonic/fermionic QNMs and GFs of the Kerr-like black hole. The paper explores the impact of the Lorentz symmetry-breaking (LSB) parameter $\alpha$ on the quasinormal modes with the help of the 6th-order WKB method \cite{Jha:2023vhn} and the study reveals that the emission frequency and decay rate initially decreases with $\alpha$ and then grows up. In \cite{Boonserm:2023oyt}, the authors investigate the greybody factor from the massive scalar field in both the asymptotically $dS$ and the $AdS$ spacetime using the WKB and the rigorous bound methods and found that the greybody factor depends on the shape of the potential as found in quantum mechanics. The paper \cite{Konoplya:2020cbv} shows the  calculation of  grey-body factor for Dirac, electromagnetic and gravitational fields and estimaion of the intensity of Hawking radiation and lifetime for asymptotically flat black holes in this theory. The study reveals that positive coupling constant leads to much smaller evaporation rate and longer life-time of a black hole, while the negative one enhances Hawking radiation. The grey-body factors for electromagnetic and Dirac fields are smaller for larger values of the coupling constant.

		In case of BHs, grey body factors measure the transmission probability of the Hawking radiation. While, in case of WHs as there is no horizon there is no intrinsic Hawking radiation but grey body factors in WHs reflect how waves traverse or are reflected by the throat's effective potential barrier. QNMs on the other hand describe how BH ``ringdown" (dying echo of a cosmic event) after perturbations \cite{Silva:2022srr}, \cite{Bae:2023sww}. QNMs depend on mass, charge and spin \cite{Zhao:2022lrl}, \cite{Konoplya:2011qq}.  However, WHs in $f(Q)$ gravity has no damping from an absorbing surface. Further, it is seen how non-metricity affects the magnitude of the oscillation frequencies. QNMs in case of BHs are strongly damped due to horizon absorption, while in case of $f(Q)$ WHs there is no damping. So, we might not experience a ``ringdown" in case of $f(Q)$ WHs. Further, the expression for grey body factor shows that as $Re(\omega_{0})$ is very very small (caused by $Q<<0$), the grey body factor grows larger indicating more transmission rate. This is the distinct signature of non-metricity in the present context. 
	
	Therefore, the study of QNMs and grey body factors bridges theoretical predictions with observational aspects in gravity wave (GW) astronomy. The effective potential analysis for WH1 confirmed the presence of stable QNMs, and the transmission coefficients indicated potential observational signatures in GW detection. The correspondence between QNMs and grey body factors hints that future GW observations could potentially serve as a tool to distinguish WHs from BHs, offering new avenues on exotic astrophysical objects.
	
	Overall, this study strengthens the viability of $f(Q)$ gravity as a compelling alternative to General Relativity (GR) in explaining exotic astrophysical structures like  the WHs. The obtained results provide meaningful insights into the theoretical modeling and observational signatures of WHs, motivating further research into their detection and implications for fundamental physics. The theoretical rigor of the work gives us the insight that unique signature of non-metricity would be in principle able to distinguish WHs from BHs in future gravity wave or electromagnetic observations using instruments with high precision.
	\section*{Acknowledgment}
	The authors thank the anonymous learned referee for valuable and insightful comments that enhanced that quality and visibility of the work. M.C thanks Department of Mathematics, Techno India University, West Bengal and S.C thanks Brainware University, West Bengal for giving research facilities.
	
\end{document}